\documentclass[pdftex,onecolumn,epjc3]{svjour3}          % twocolumn
\usepackage[a4paper,left=2.5cm,right=2.5cm,top=3cm, bottom=2.0cm]{geometry} %inserido 

\newcommand{\pd}[2]{\frac{\partial #1}{\partial #2}} 
\newcommand{\pdd}[2]{\frac{\partial^2 #1}{\partial #2^2}} 
\newcommand{\abs}[1]{\left| #1 \right|} % for absolute value

\RequirePackage[T1]{fontenc}

\smartqed  % flush right qed marks, e.g. at end of proof

\RequirePackage{graphicx}
\RequirePackage{mathptmx}      % use Times fonts if available on your TeX system
\RequirePackage{flushend}
\RequirePackage[numbers,sort&compress]{natbib}
\RequirePackage[colorlinks,citecolor=blue,urlcolor=blue,linkcolor=blue]{hyperref}

\journalname{The European Physical Journal D}

\begin{document}

\title{Coulomb Correlation and Information Entropies in Confined Helium-Like Atoms}


\author{Wallas Santos Nascimento\thanksref{e1,addr1}
        \and
        Marcos Melo de Almeida\thanksref{e2,addr1} %etc.
\and
        Frederico Vasconcellos Prudente\thanksref{e3,addr1}
}

%\thankstext[$\star$]{t1}{Thanks to the title}
\thankstext{e1}{e-mail: wallassantos@gmail.com}
\thankstext{e2}{e-mail: marcosma@ufba.br}
\thankstext{e3}{e-mail: prudente@ufba.br}

\institute{Instituto de F\'{\i}sica, Universidade Federal da Bahia, 40170-115, Salvador, Bahia, Brazil.\label{addr1}
          %\and
          %Second Address, Street, City, Country\label{addr2}
          %\and
          %\emph{Present Address:} Street, City, Country\label{addr3}
}

\date{Version accepted for publication in Eur. Phys. J. D (2021). \href{https://doi.org/10.1140/epjd/s10053-021-00177-6}{DOI: 10.1140/epjd/s10053-021-00177-6}}

\abstractdc{ The present work studies aspects of the electronic correlation in confined H$^{-}$, He and Li$^+$ atoms in their ground states using the informational entropies. In this way, different variational wavefunctions are employed in order of better take account of Coulomb correlation. The obtained values for the $S_r$, $S_p$ and $S_t$ entropies are sensitive in relation to Coulomb correlation effects. In the strong confinement regime, the effects of the Coulomb correlation are negligible and the employment of the models of independent particle and two non-interacting electrons confined by a impenetrable spherical cage gains importance in this regime. Lastly, energy values are obtained in good agreement with the results available in the literature.
}

\maketitle

\section{Introduction}

The problem of many bodies is best characterized when the constituent particles of the system interact simultaneously, that is, the particles have a correlated behavior. Specifically, the electronic interactions in atoms or molecules globally generate so-called correlation effects~\cite{correlacao_livro,statistical_correlation}. 

The phenomenon of electron correlation rises when working with models that go beyond the approximation of the independent particle model, being established by means of two essential origins. The first one is denominated Fermi correlation and results from the hypothesis of indistinguishability of the fermionic particles, producing an exchange force between them. The second one is known as Coulomb correlation based on the Coulomb repulsion between the electrons~\cite{tew,fermicoulombholes}.

Usually, the electronic correlation is studied as a guide in an energetic analysis of the systems. In this sense, the so-called correlation energy~\cite{correlacao_original,baskerville}, defined as the difference between the exact and limit Hartree-Fock energies, has received significant attention. The correlation energy, although small related to total energy, is very relevant in determination of chemical bonding energies. Futhermore, other energy analysis procedures can be defined~\cite{correlacao_dif3, correlacao_dif4}.

A mathematical theory of communication or information theory arose with report of Claude Shannon in 1948. A fundamental concept in this field is the information or Shannon entropy. This quantity is employed to measure the amount of information generated by a information source in the process of choosing a message from a possible group of them~\cite{rioul, shannonoriginal}. Information entropy is successfully used in the study of the most varied phenomena in diverse fields of expertise~\cite{WANG2019429, LIU20181170}.

From the connection between information and quantum theories, it emerges the informational entropies in position, $S_r$, and in momentum, $S_p$, spaces, and the entropy sum $S_t$ by adding $S_r$ and $S_p$. The study of quantum systems in information theory context already has a series of published works, for instance, analyses involving the one-dimensional systems~\cite{robinquantumwell,wallas_marcos_fred, oscillatorinsymmetricandasymmetric,quantumuncertainties,sun} and confined
hydrogen atom~\cite{estanon,mukherjee-roy2018,hydrogenatomconfineddifferentpotentials,wallas_fred_internacional,salazar,dehesa}. Studies on the free or confined helium-like atoms also have received attention in the informational context~\cite{hao2019,nassar4,NASSER20173892,entropiaatomosconfinados,majumdar, shannon_dois_eletrons2, shannon_dois_eletrons3,nasser2,MARTINEZFLORES,martinezflores2}.

Spatially confined quantum systems are not a recent problem in physics and have their initial studies produced in the early 1930s~\cite{michael, fock, darwin}. Quantum mechanical properties of confined systems are influenced by properties of confinement cage, for instance, the energy spectra and the polarizability experience significant changes~\cite{dipolo, dipoloenergiaelso}. With the increment of computational processing capacity and development of modern techniques for the effective confinement, this topic remains to be of interest~\cite{FLORESRIVEROS20086175,PhysRevA.93.022512,indicacao1,connerade2020,saha-jose2020,hartreefockstudy,prasad2,maniero2020,batel2021,fred_marcilio_livro,garza,softconfinementpotentials}. In previous researches, we use the entropy sum to investigate the regions of the confined harmonic oscillator~\cite{wallas_fred_educacional} and confined hydrogenic-like atoms~\cite{wallas_fred_internacional} where the confinement effects has major impact.

Despite of the studies using energy analysis to measure electronic correlation, studies based on informational entropies are found in literature~\cite{saha_jose,correlacao, shannon_correlacao3,shannon_correlacao4,  entropy_of_fractional_occupation_probability, informacao_mutua_e_correlaca_momomento}. The investigation in free quantum systems, \emph{i.e.}, non spatially confined, shows promising results. For the lithium~\cite{correlacao5} and beryllium~\cite{correlacao6} atoms, and their ions, the $S_t$ values increase and the $S_r$ and $S_p$ values increase and decrease, respectively, using configuration interaction singles and doubles excitations instead of Hartree-Fock wavefunctions~\cite{shannon_correlacao2}.  

The goal of this work is to discuss the Coulomb correlation effects on the systems by means of their wavefunctions, employing informational entropies. The systems of interest are spatially confined atoms containing two electrons, namely: negative hydrogen ion (H$^{-}$), helium (He) and ionized lithium (Li$^+$). This paper is organized as follows: in sect.~\ref{Theoretical background} we present the informational concepts employed and we analyze the physical systems of interest together with the trial wavefunctions, in sect.~\ref{Results} we discuss the obtained results and, finally, in sect.~\ref{conclusions} we summarize the central aspects of our investigation. In Appendix A we present the $2e^{-}$ system.

\section{Theoretical background}{\label{Theoretical background}}

In this section, we present the concepts and methods applied in our research. In subsect.~\ref{Information entropies} we define the informational quantities $S_r$, $S_p$ and $S_t$. In subsect.~\ref{Systems of interest} we examine the confined helium-like atoms, as being the physical system of interest in this work. Furthermore, we define the three trial wavefunctions that we use to describe the confined H$^{-}$, He and Li$^+$ atoms in their ground energy states in order of better accuracy in the Coulomb correlation description.

\subsection{Information entropies}{\label{Information entropies}}

The expression, originally introduced by Shannon, employed to measure the amount of information generated by a continuous information source is~\cite{shannonoriginal}
\begin{equation}
S(p(\alpha))=  - \int\limits_{-\infty}^{\infty} d\alpha \ p(\alpha) \log_2 p(\alpha)  \ , 
\label{shannoncontinuo}
\end{equation} 
where $p(\alpha)$ is a probability density in function of continuous $\alpha$ variable, constrained by normalization condition, $\int\limits_{-\infty}^{\infty} d\alpha \ p(\alpha) =1$, and non-negative condition, $p(\alpha) \geq 0$, $\forall \, \alpha$. Note that in the original Shannon's work the entropy values for continuous distributions may be negative (see pg. 631 in Ref.~\cite{shannonoriginal}).

Adopting on the expression~(\ref{shannoncontinuo}) the two-electron probability densities in position and momentum spaces, respectively $\rho(\vec{r}_1, \vec{r}_2)$ and $\gamma(\vec{p}_1, \vec{p}_2)$, arising from quantum theory, and using the fundamental physical constants Bohr radius, $a_0$, and reduced Planck constant, $\hbar$, we have the information entropies in position and momentum spaces, respectively, defined as~\cite{wallas_fred_internacional} 
\begin{equation}
S_r = -\int \int d\vec{r}_1\  d\vec{r}_2 \ \rho(\vec{r}_1, \vec{r}_2) \ln \left( {a_0}^6 \ \rho(\vec{r}_1, \vec{r}_2) \right)
\label{entropia_posicao}
\end{equation}
and
\begin{equation}
S_p = -\int \int   d\vec{p}_1 \ d\vec{p}_2  \ \gamma(\vec{p}_1, \vec{p}_2) \ln \left( {\left( \frac{\hbar}{a_0} \right)^6} \ \gamma(\vec{p}_1, \vec{p}_2)  \right) \ .
\label{entropia_momento}
\end{equation}
The $\rho(\vec{r}_1, \vec{r}_2)$ and $\gamma(\vec{p}_1, \vec{p}_2)$ probability densities are normalized to unity and are written in terms of two-electron wavefunctions, $\rho(\vec{r}_1, \vec{r}_2)= {\arrowvert \psi(\vec{r}_1, \vec{r}_2) \arrowvert}^2 $ and $\gamma(\vec{p}_1, \vec{p}_2)={\arrowvert \widetilde{\psi}(\vec{p}_1, \vec{p}_2)\arrowvert}^2$. The momentum space wavefunction $\widetilde{\psi}(\vec{p}_1, \vec{p}_2)$ is obtained by Fourier transform in the position space of wavefunction $\psi(\vec{r}_1, \vec{r}_2)$. 

The $S_t$ quantity is obtained from entropy sum, $S_r + S_p$, as defined below. From the entropy sum we can still derive the entropy uncertainty relation that, written in terms of $\hbar$, is given by~\cite{wallas_fred_internacional}
%\begin{footnotesize}
\hspace*{-1.5cm}\begin{eqnarray}
S_t & = & S_r+S_p \nonumber \\
& = & - \int \cdots \int d\vec{r}_1 \  d\vec{r}_2  \ d\vec{p}_1 \ d\vec{p}_2 \  \rho(\vec{r}_1, \vec{r}_2)  \ \gamma(\vec{p}_1, \vec{p}_2) \ \ln\left( \hbar^6 \  \rho(\vec{r}_1, \vec{r}_2)  \ \gamma(\vec{p}_1, \vec{p}_2) \right) \nonumber \\
&\geq& 6 \ (1+\ln\pi) \ .
\label{Str}
\end{eqnarray}
%\end{footnotesize} 
From this entropy uncertainty relation we can obtain the Kennard's uncertainty relation~\cite{relacaodeincertezainformacao}. The Eqs.~(\ref{entropia_posicao}), (\ref{entropia_momento}) and (\ref{Str}) are dimensionally adequate and independent of the employed unit system. 

The $S_r$ and $S_p$ entropies are a type of dispersion or spread measures that are calculated without taking into account reference points in the probability distributions as Kennard's uncertainty relation (see, for example, Ref.~\cite{wallas_marcos_fred} and references therein). In this sense, the entropic quantities are measures of uncertainty, localization or delocalization, of the wavefunction in the space~\cite{localizationdelocalization,wallas_marcos_fred}. 
 
\subsection{Systems of interest}{\label{Systems of interest}}

The confined two-electrons (or the confined helium-like) atoms are interesting and simple systems to explore the different aspects of confinement. The complete non-relativistic Hamiltonian in atomic units for the confined helium-like atoms at the center of an impenetrable spherical cavity is given by
\begin{equation}
\hat{H} = \sum_{i=1}^2  \hat{h}(\vec{r}_i) + V_{ee}(\vec{r}_1,\vec{r}_2)  \ ,  
\label{hamiltonianohelio}
\end{equation}
where $\hat{h}(\vec{r}_i)$ is the confined hydrogenic-type Hamiltonian of the $i$-th electron,
\begin{equation}
    \hat{h}(\vec{r}_i)= -\frac{1}{2} \nabla_i^2 - \frac{Z}{r_i} + V_c (r_i) \ ,
    \label{hi}
\end{equation}
while $V_{ee}$ corresponds to electrostatic electron-electron repulsion,
\begin{equation}
    V_{ee}(\vec{r}_1,\vec{r}_2) = 1 / r_{12} \ .
    \label{vee}
\end{equation}
In above equations, $r_i$ represents the distance from the nucleus to the $i$-th electron, $r_{12}$ is the relative distance between the electrons, and the atomic number, $Z$~=~1, 2 and 3, defines the negative hydrogen ion (H$^{-}$), helium (He) and the ionized lithium (Li$^+$), respectively. In present work we assume the spherical confining potential as
\begin{equation}
V_c (r_i)=  \left \{ \begin{array}{ccc}
0 & \mathrm{for} &  0 < r_i < r_c  \\
\infty & \mathrm{for} &  r_i \geq r_c 
\end{array}\right. ,
\label{potencialhelio}
\end{equation}  
being $r_c $ the confinement radius.

Since this is a fermionic system, the electronic total wavefunction has to be antisymmetric by exchange of both electrons due to Pauli exclusion principle. This assumption includes Fermi correlation to the problem. As the Hamiltonian of the problem is spin independent, the product of spatial and spin parts constitute the total wavefunction. Assuming that the spin part is antisymmetrical for the electronic ground state (a singlet electronic state), the spatial part of wavefunction must be symmetrical by exchange of two electrons. 

In present work we select three different trial spatial wavefunctions to describe confined helium-like atoms in their ground states  in order of better take account of Coulomb correlation. The spin part will be omitted because the Hamiltonian is spin independent.

Our starting point is an exact function for the independent particle model $\varphi_{1}$. The effect of the electrostatic repulsion [Eq.~(\ref{vee})] is regarded as a perturbation in computation of energies. In this independent electron model, the trial wavefunction is given by
\begin{equation}
\varphi_{1} (r_1, r_2) = \ A \ e^{\alpha (r_1+r_2)} \ \Omega(r_1, r_2) \ ,
\label{heliosc}
\end{equation}  
where the orbital exponent $\alpha$ is adjusted to minimizing the electronic energy from the non-interacting particles Hamiltonian $\hat{H}_0= \hat{h}(\vec{r}_1) + \hat{h}(\vec{r}_2)$. The total ground state energy is found by introducing the electron-electron interaction through the first order perturbation theory. In this case, the one-electron probability density is exactly the same of the confined hydrogenic-type system. In Eq.~(\ref{heliosc}) and the following, $\Omega(r_1, r_2)$ is the cutoff function which provides the appropriate boundary conditions.

On the other hand, the complete Hamiltonian [Eq.~(\ref{hamiltonianohelio})] is considered in the two other models. The difference between them is how the correlation effect is introduced into the trial wavefunction.  In the second model the trial wavefunction $\varphi_{2}$ has a similar analytical expression to the first model, that is,
\begin{equation}
\varphi_{2} (r_1, r_2)  = B \ e^{\beta (r_1+r_2)}\ \Omega(r_1, r_2) \ ,
\label{helioic}
\end{equation}
but the exponent orbitals $\beta$ is now obtained variationally by minimizing the total electronic energy functional. In this case, an electron interacts with the average field produced by other electron. The apparent effect of each electron on the movement of the other is to reduce the nuclear charge, introducing a change in the probability density when compared to the first model.

The third model considered here consists of introducing an explicit correlation factor between the two electrons in the trial wavefunction, as follows: 
\begin{equation}
\varphi_{3} (r_1, r_2)  = C \ e^{\lambda (r_1+r_2)} \ \gamma(r_{12}) \ \Omega(r_1, r_2) \ ,
\label{heliomc}
\end{equation}
where the correlation factor $\gamma(r_{12})$ is given by
\begin{equation}
\gamma(r_{12}) = \left(1 + \sigma \ r_{12} \right) \ . 
\end{equation}
The parameters $\lambda$ and $\sigma$ are determined by the variational method. The inclusion of a new variational parameter in $\varphi_{3}$ function makes it more flexible and we expected a wavefunction that better describes the system. In this way, part of correlation energy is recovered by $\varphi_{3}$ function. 

In $\varphi_{1}$, $\varphi_{2}$ and $\varphi_{3}$ functions, the normalization constants are $A$, $B$, and $C$, and the cutoff function which provides the appropriate boundary conditions is
\begin{equation}
\Omega (r_1, r_2) = \left[1- \left( \frac {r_1} {r_c} \right) \right] \left[ 1- \left(\frac {r_2} {r_c} \right) \right] \ .
\end{equation}

The Eq.~(\ref{heliomc}) is a correlated wavefunction by $\gamma(r_{12})$ factor. The inclusion of this factor introduces considerable complications in problem solution. To solve it, we employ the Hylleraas coordinates~\cite{hylleraaslivro,hylleraasquimicanova,integraisconfinadashylleraas,novasexpressoeshylleraas}, which are written as
\begin{equation}
s \equiv r_2 + r_1 \ , \ \ t  \equiv r_2 - r_1 \ \ \mathrm{and} \ \ u \equiv \abs{\vec{r}_1 - \vec{r}_2} \ .
\label{eq:subeq1} 
\end{equation}
In these coordinates the Hamiltonian of the system is
\begin{eqnarray}
\mathrm{\widehat{H}}&=&- \left( \pdd{}{s}+\pdd{}{t}+\pdd{}{u} \right) -\frac{4s}{s^2-t^2}\pd{}{s}+\frac{4t}{s^2-t^2}\pd{}{t}-\frac{2}{u}\pd{}{u}- \nonumber\\
 & & -2\frac{s}{u} \left(\frac{u^2-t^2}{s^2-t^2} \right) \frac{\partial^2}{\partial s \partial u}- 2\frac{t}{u} \left(\frac{u^2-s^2}{s^2-t^2} \right) \frac{\partial^2}{\partial t \partial u}-4Z\frac{s}{s^2-t^2}+\frac{1}{u} \ .
\end{eqnarray} 
The correlated wavefunction is written as
\begin{equation}
\varphi_{3}(s, t, u) = G \ e^{\sigma s} \left(1+b \ u \right) \Upsilon(s, t) \ ,
\end{equation}
the normalization constant is $G$. For the cutoff function we have
\begin{equation}
\Upsilon(s, t) = \left[ r_c - \left(\frac{s - t}{2}\right) \right] \left[ r_c - \left(\frac{s + t}{2}\right) \right] \ .
\end{equation}
The volume element in terms of Hylleraas coordinates is
\begin{equation}
dv= 2 \pi^2 (s^2 - t^2) \ u \ ds \ dt \ du \ ,
\label{dvdsdtdu}
\end{equation}
with the limits of integration being 
\begin{eqnarray}
\int \ ds \ du \ dt \ &=&  \int_{0}^{r_c} ds  \int_{0}^{s} du  \int_{0}^{u}dt + \int_{r_c}^{2r_c} ds  \int_{0}^{2r_c-s} du  \int_{0}^{u}dt +    \label{intdsdudtconfinado}    \nonumber \\
      & & + \int_{r_c}^{2r_c} ds  \int_{2r_c-s}^{s} du  \int_{0}^{2r_c-s}dt \ .   
\end{eqnarray}
The Hylleraas coordinates are considered to be coordinate systems of two electrons.

\section{Results and discussions}{\label{Results}}

In this section, we present and discuss our results for the confined H$^{-}$, He and Li$^+$ atoms (ground states energy) under various confinement conditions. In subsect.~\ref{energy analysis} we consider the results of expectation value of the energy. In subsect.~\ref{Information entropiesr}, we determine the informational entropies $S_r$ and $S_p$ using the Eqs.~(\ref{entropia_posicao}) and~(\ref{entropia_momento}); and, once get these values, we find the entropy sum $S_t$. In subsect.~\ref{Strong confinement} we produce an informational study in the strong confinement regime.

We use the softwares Maple13 and QtiPlot to perform the calculations and plot the graphs, respectively. In particular, the numerical calculations, including the numerical integrations, were performed with an accuracy of at least 10 digits. Thus, even with an eventual propagation of numerical errors, we estimate that our results have an accuracy of at least 6 digits, so that the values presented in the tables are accurate within the significant figures presented.

We use the atomic units (a.u.) system. This system uses the electronic mass, $m_e$, the elementary charge $e$ of the electron, the constant of electrostatic force, $1/4 \pi \varepsilon_0$, and the reduced Planck constant, $\hbar$, as standard units of their respective quantities. Here, we consider $m = m_e$ for all the calculations.    

\subsection{Energy analysis} {\label{energy analysis}}

\begin{figure}[ht] 
\centering
\includegraphics[scale=0.65]{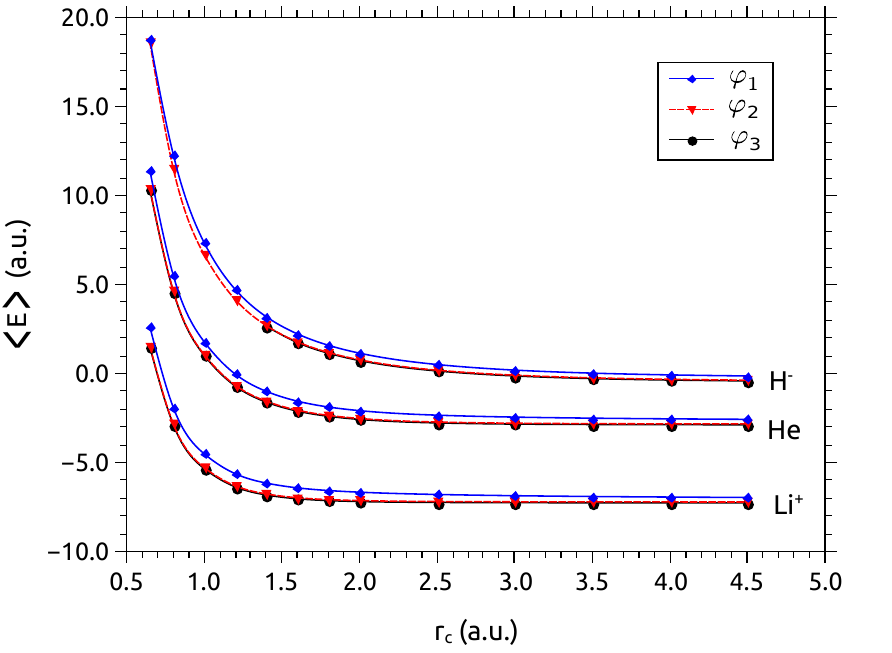}
\caption{Values of  $\langle E \rangle$ as a function of $r_c$ for the confined H$^{-}$, He and Li$^+$ atoms in their ground states.}
\label{energiasduasparticulasfig}
\end{figure}

The expectation value of the energy, $\langle E \rangle$, for the confined H$^{-}$, He and Li$^+$ atoms in their ground states for three wavefunctions $\varphi_{1}$, $\varphi_{2}$ and $\varphi_{3}$, as a function of $r_c$, are displayed in Fig.~\ref{energiasduasparticulasfig} and organized in Table~S.1 of the Supporting Information. We determine the $\langle E \rangle$ values of $\varphi_{1}$ adding the values from model independent particles, $\langle E^0 \rangle$, plus a first correction, $\langle E^1 \rangle $, established by expectation value of the $V_{ee}$ term. Adjusting the variational parameter in $\varphi_{2}$, with the complete Hamiltonian~(\ref{hamiltonianohelio}), we obtain better energy values than $\varphi_{1}$. Besides, using the $\gamma(r_{12})$ term in $\varphi_{3}$, we get the best results in this work, recovering a part of the correlation energy. We detect that the energy values improve in order of consideration of the Coulomb correlation. We can note, from Fig.~\ref{energiasduasparticulasfig}, that $\langle E \rangle$ values tend to the values of the unconfined system (free system) when $r_c$ goes to infinity. In another situation, the curves of $\langle E \rangle$ blow up when the value of $r_c$ becomes smaller.

The $\langle E \rangle$ values taken by $\varphi_{3}$ for all three isoelectronic members are very close and, occasionally, are lower energies than the results of the Ref.~\cite{entropiaatomosconfinados} that makes its analysis from density functional theory. For He the energy values provided by $\varphi_{2}$ and $\varphi_{3}$ have a good agreement with the results presented by the Ref.~\cite{gimarccorr}, which also performs a variational calculation. Such comparisons indicate that the values of energy obtained here are compatible with the results published in the literature.

\begin{figure}[!ht]
\centering
\includegraphics[scale=0.6]{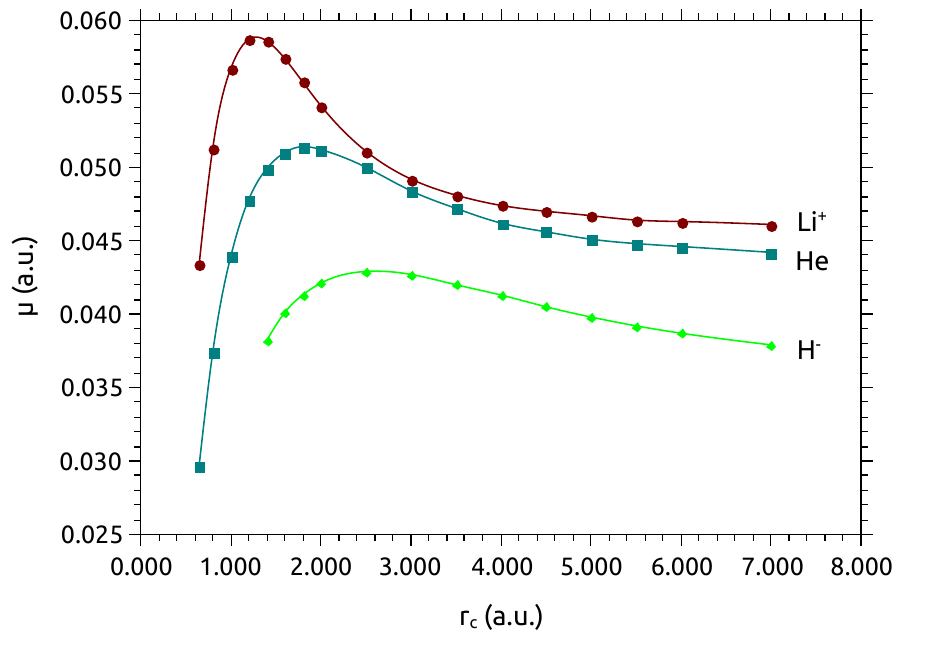}
\caption{Values of  $\mu$ as a function of $r_c$ for the confined H$^{-}$, He and Li$^+$ atoms in their ground states.}
\label{u2g}
\end{figure}

The investigation of the correlation effects under different confinement conditions, can be done by difference between the expectation value of the energies $\langle E\left[ \varphi_{2} \right] \rangle $ and $\langle E \left[ \varphi_{3} \right] \rangle $, determined by $\varphi_{2}$ and $\varphi_{3}$, respectively, as follows:
\begin{equation}
\mu = \langle E\left[ \varphi_{2} \right] \rangle - \langle E \left[ \varphi_{3} \right] \rangle  \ .
\end{equation}
The quantity $\mu$ is displayed in Fig.~\ref{u2g} as a function of $r_c$. Note that in the atoms studied here the $\mu$ quantity increases with $Z$ increment, and that each $\mu$ associated with a specific confined atomic system has a peak, whose maximum value and position is: the H$^{-}$, $\mu^{max}\approx0.0429$ at $r_c\approx2.5603$; He, $\mu^{max}\approx0.0514$ at $r_c\approx1.8132$; and, finally, for the Li$^+$, $\mu^{max}\approx0.0589$ at $r_c \approx1.2765$. For larger $r_c$, the $\mu$ values tend to their free atom values. Moreover, when $r_c$ goes to zero, the $\mu$ values decrease, indicating that the energy values of $\varphi_{3}$ approaches of $\varphi_{2}$. In the scenario, the correlation factor $\gamma(r_{12})$ loses importance on the characterizing of the wavefunction. The $\mu$ values are close to true correlation energy ($E_{HF} - E_{exact}$) for the unconfined or confined two-electrons atoms~\cite{gimarccorr}.

One thing to note is about the H$^-$ anion. The energy values produced by $\varphi_{1}$, $\varphi_{2}$ and $\varphi_{3}$ could represent the H$^{-}$ as an unstable arrangement. However, the imposed confinement barriers avoid that one of the electrons detaches from the atom. In the free H$^{-}$ the electrons are loosely attached to the nucleus, so its structure could be energetically appropriate to a dissociation in a neutral hydrogen atom plus an ejected electron. These particularities of the H$^{-}$ can to difficult the study of this ion through the variational method and to produce some numerical instability in the determination of physical quantities of interest in regions of rigorous confinement. Nevertheless, investigations demonstrate that a unique bound state is observed when using more elaborate variational wavefunctions~\cite{ionnegativohidrogenio,hill,atomose,LeSech}.

\subsection{Information entropies}{\label{Information entropiesr}}

\begin{figure}[h!]
\centering
\includegraphics[scale=1.52]{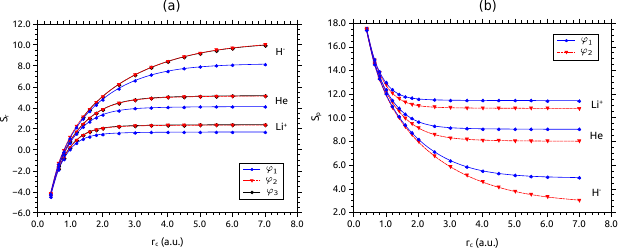}
\caption{Values of $S_r$ and $S_p$ as a function of $r_c$ for the confined H$^{-}$, He and Li$^+$ atoms in their ground states.}
\label{entropias}
\end{figure}

The $S_r$ values as a function of $r_c$ for the confined H$^{-}$, He and Li$^+$  atoms in their ground states are showed in Fig.~\ref{entropias} and organized in Table~\ref{entropiap} and Table~S2 of the Supporting Information. The first observation is that, for the same $r_c$, the informational entropies determined using the trial $\varphi_1$ function are exactly twice the informational entropies of equivalent hydrogenic-like systems obtained previously by us in Ref.~\cite{wallas_fred_internacional}. For this comparison, we consider the data calculated using the trial $\phi^c_{11}$ function of such publication. The reason for this agreement comes from the fact that its exponential $\alpha$  parameter was obtained disregarding the electron-electron repulsion term. Besides, when $r_c$ goes to infinity the $S_r$ values tend to the $S_r$ values of the unconfined systems presented in Ref.~\cite{NASSER20173892}, indicating the validity of our results. It should be noted that, in their work, Nasser {\it et al.} used wavefunctions with the same functional forms of $\varphi_2$ and $\varphi_3$, excluding the cutoff function, to treat the free two-electron helium-like atoms.

\begin{table}[!ht]
\caption{Values of $S_r$ as a function of $r_c$ for the H$^{-}$, He and Li$^{+}$ atoms in their ground states using the functions $\varphi_{1}$, $\varphi_{2}$ and $\varphi_{3}$. Values of $r_c$ in atomic units system. The literature references are given as superscript.} 
\label{entropiap}
\centering
\begin{tabular}{cccccccccccccccccc}
\\ \hline
%& & \multicolumn{12}{c}{$S_r$}
& \vline &	\multicolumn{3}{c}{H$^{-}$}  & \vline & \multicolumn{3}{c}{ He } & \vline & \multicolumn{3}{c}{Li$^{+}$}  
\\ \hline
$r_c$ & \vline &  $\varphi_{1}$	& $\varphi_{2}$  & $\varphi_{3}$ & \vline &  $\varphi_{1}$ 	& $\varphi_{2}$  & $\varphi_{3}$	& \vline & $\varphi_{1}$	 & $\varphi_{2}$ 	&  $\varphi_{3}$
\\ \hline
1.0000& \vline&	1.0823&	1.1760&	---------& \vline&	0.6345&	0.7734&	0.7779& \vline&	-0.0029&	0.1973&	0.2071
\\ \hline
2.0000& \vline&	4.7933&	5.0561&	5.0543& \vline&	3.3254&	3.8232&	3.8400& \vline&	1.4804&	2.0257&	2.0578
\\ \hline
4.0000& \vline&	7.4843&	8.4092&	8.3985& \vline&	4.0407&	4.9813&	5.0173& \vline&	1.6670&	2.3137&	2.3514
\\ \hline
6.0000& \vline&	8.0720&	9.7039&	9.6810& \vline&	4.0998&	5.0960&	5.1336& \vline&	1.6860&	2.3413&	2.3793
\\ \hline
10.0000& \vline&	8.2406&	10.3519&	10.3056& \vline&	4.1212&	5.1345&	5.1724& \vline&	1.6939&	2.3525&	2.3906
\\ \hline
free &\vline& ---------	&	10.538$^{\mathrm{a}}$ &	10.479$^{ \mathrm{a}}$ & \vline&	---------&	5.1500$^{ \mathrm{a}}$&	5.1881$^{\mathrm{a}}$& \vline&	---------&	2.3577$^{ \mathrm{a}}$&	2.3959$^{ \mathrm{a}}$
\\ \hline
&\multicolumn{12}{l}{$^{ \mathrm{a}}$\begin{small}Ref.~\cite{NASSER20173892}                                         \end{small}}  \\
\end{tabular}
\end{table}

From graph (a) of the Fig.~\ref{entropias}, we see that the difference between $S_r$ values determined by $\varphi_1$ and $\varphi_2$ are higher than $\varphi_{2}$ and $\varphi_{3}$. The $S_r$ curves obtained by  $\varphi_{2}$ and $\varphi_{3}$ are close to producing an overlap in such figure. Moreover, when the confinement increases the $S_r$ curves tend to the negative values. In Table~\ref{entropiap} and Table~S2 of the Supporting Information, the results of $S_r$ for the He and Li$^+$ atoms increase using the functions $\varphi_{1}$, $\varphi_{2}$ and $\varphi_{3}$, in this order. For the H$^{-}$, the $S_r$ values determined by $\varphi_{1}$ are less than those employing $\varphi_{2}$, whereas the calculated by $\varphi_{2}$ are greater than those provided by $\varphi_{3}$. This phenomenon occurs, in H$^{-}$, because the use of $\gamma(r_{12})$ induces the decoupling of an electron from atom, while the other get closer of the nucleus. This effect produces a more localized probability density in position space and, consequently, a lower value for $S_r$ using $\varphi_3$.

Also based on the graph (a) of the Fig.~\ref{entropias} note that the values of $S_r$ can be negative. These values correspond to regions where the probability densities are highly localized. This fact can also be observed in one-dimensional systems~\cite{wallas_marcos_fred}, in the spherically confined hydrogen atom~\cite{wallas_fred_internacional} and systems with static screened Coulomb potential~\cite{screened_Coulomb_potential}. This result has a simple explanation in the quantum context~\cite{AQUINO20132062}: when $r_c$ is reasonably small, the probability density becomes large and $a_0^6\rho(\vec{r}_1, \vec{r}_2) > 1$. In this situation, $-\rho(\vec{r}_1, \vec{r}_2) \ln (a_0^6\rho(\vec{r}_1, \vec{r}_2)) < 0$ and, then, $S_r$ may be negative.

We visualize the performance of the $S_p$ versus $r_c$ curves for $\varphi_{1}$ and $\varphi_{2}$ in the graph (b) of the  Fig.~\ref{entropias} and we organize the numerical values in Table~\ref{entropiam} and Table~S3 of the Supporting Information. The incorporation of the Coulomb correlation on the wavefunction $\varphi_{2}$ causes a reduction in the $S_p$ values. When $r_c$ goes to infinity the $S_p$ values pass to constant values. In another situation, when the confinement increases, an upward inflection of the curves is identified. The integral calculations of the information entropy on momentum space showed numerical instability and were truncated to a given value of the integration limit. Taking that into account, we guarantee the accuracy of the results obtained of $S_p$ using the functions $\varphi_{1}$ and $\varphi_{2}$. Studies involving Fourier transforms in correlated wavefunctions have been started~\cite{saavedra1996}, allowing us in the future to provide the $S_p$ values for the function~$\varphi_{3}$.

\begin{table}[!ht]
\caption{Values of $S_p$ as a function of $r_c$ for the H$^{-}$, He and Li$^{+}$ atoms in their ground states using the functions $\varphi_{1}$ and $\varphi_{2}$. Values of $r_c$ in atomic units system.}
\label{entropiam}
\centering
\begin{tabular}{ccccccccccccccccc}
\\ \hline
& \vline &	\multicolumn{2}{c}{ H$^{-}$}	  & \vline & 	\multicolumn{2}{c}{ He}	   &	 \vline & \multicolumn{2}{c}{ Li$^{+}$} 
\\ \hline
$r_c$ & \vline &  $\varphi_{1}$	& $\varphi_{2}$	  & \vline & $\varphi_{1}$	& $\varphi_{2}$  	& \vline & $\varphi_{1}$	 & $\varphi_{2}$	
\\ \hline
1.0000& \vline &	12.0076&	11.9496& \vline &	12.3575&	12.2374& \vline &	12.9705&	12.7709
\\ \hline
2.0000& \vline &	8.1986&	7.9796& \vline &	9.6665&	9.1507& \vline &	11.5832&	11.0197
\\ \hline
4.0000& \vline &	5.5076&	4.5581& \vline &	9.0582&	8.0998& \vline &	11.4537&	10.8032
\\ \hline
6.0000& \vline &	4.9915&	3.2971& \vline &	9.0209&	8.0183& \vline &	11.4424&	10.7857
\\ \hline
10.0000& \vline &	4.8731&	2.7320& \vline &	9.0081&	7.9931& \vline &	11.4377&	10.7787
\\ \hline
\end{tabular}
\end{table}

The Ref.~\cite{shannon_dois_eletrons} helps us to understand the behavior of $S_r$ and $S_p$ with the $Z$ variation. The increase of the atomic number produces lower results of the expectation values of $r_1$ and $r_{12}$, making more localized systems, consequently the $S_r$ values are reduced with $Z$ increment in Table~\ref{entropiap} and Table~S2 of the Supporting Information. On the other hand, the reduction of the atomic number makes weaker interactions. The movement of particles is less chaotic, then the $S_p$ values decrease with $Z$ reduction in Table~\ref{entropiam} and Table~S3 of the Supporting Information.

\begin{figure}[!ht]
\centering
\includegraphics[scale=0.75]{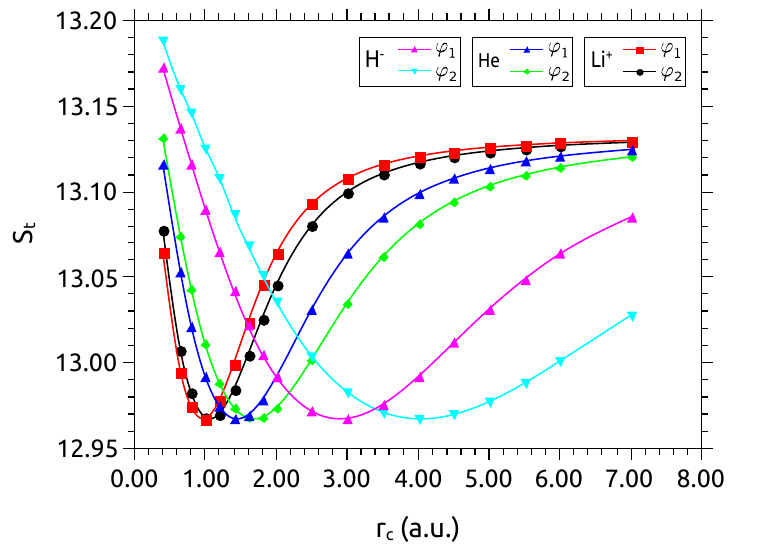}
\caption{Values of $S_t$ as a function of $r_c$ for the confined H$^{-}$, He and Li$^+$ atoms in their ground states .}
\label{somaentropica}
\end{figure}

The $S_t$ values as a function of $r_c$ for the confined helium-like atoms in their ground states are organized in Table~\ref{somaentropicad} and Table~S4 of the Supporting Information and we display in Fig.~\ref{somaentropica} the overall behavior of it. Note that the entropic uncertainty relation ($S_t \ge 12.8684$) is satisfied for all studied confined systems. Moreover, both results obtained with $\varphi_1$ and $\varphi_2$ trial wavefunctions for the three helium-like atoms show the same overall behavior. All $S_t$ curves have a minimum for intermediate values of $r_c$. In addition, for large $r_c$, the values of $S_t$ using $\varphi_1$ and $\varphi_2$, for the confined H$^{-}$, He and Li$^+$ atoms tend towards the value of 13.1332 of the free atomic system. And in the direction of small $r_c$, the entropy sums present an increase in their values. As we have observed in subsect.~\ref{Information entropies} the entropy sum gives rise to entropic uncertainty relation. In this context, at the point where $S_t$ has its lowest value ($S_t^{min}$) the system is on the situation of lower entropic uncertainty.

\begin{table}[!ht]
\caption{Values of $S_t$ as a function of $r_c$ for the H$^{-}$, He and Li$^{+}$ atoms in their ground states using the functions $\varphi_{1}$ and $\varphi_{2}$. Values of $r_c$ in atomic units system.}
\label{somaentropicad}
\centering
\begin{tabular}{ccccccccccccccccc}
\\ \hline 
& \vline &	\multicolumn{2}{c}{ H$^{-}$}	  & \vline & 	\multicolumn{2}{c}{ He}	   &	 \vline & \multicolumn{2}{c}{ Li$^{+}$} 
\\ \hline
$r_c$ & \vline &  $\varphi_{1}$	& $\varphi_{2}$	  & \vline & $\varphi_{1}$	& $\varphi_{2}$  	& \vline & $\varphi_{1}$	 & $\varphi_{2}$	
\\ \hline
1.0000& \vline &	13.0899&	13.1256& \vline &	12.9919&	13.0107& \vline &	12.9676&	12.9682
\\ \hline
2.0000& \vline &	12.9919&	13.0357& \vline &	12.9919&	12.9739& \vline &	13.0636&	13.0454
\\ \hline
4.0000& \vline &	12.9919&	12.9673& \vline &	13.0989&	13.0811& \vline &	13.1207&	13.1169
\\ \hline
6.0000& \vline &	13.0636&	13.0010& \vline &	13.1207&	13.1143& \vline &	13.1284&	13.1270
\\ \hline
10.0000& \vline &	13.1137&	13.0839& \vline &	13.1294&	13.1276& \vline &	13.1316&	13.1312
\\ \hline
\end{tabular}
\end{table}

One interesting aspect to point out is that the values of the confinement radius, $r_c^{min}$, in which $S_t^{min}$ arise, are shifted to larger values in order of better take account of Coulomb correlation in wavefunction. For the H$^{-}$ using $\varphi_{1}$ we have $S_t^{min}~=~$12.9676 in $r_c^{min}~=~$3.0010~u.a., while for $\varphi_{2}$ we have $S_t^{min}~=~$12.9673 in $r_c^{min}~=~$4.0000~u.a.. For the He, we have for $\varphi_{1}$  the value of $S_t^{min}~=~$12.9675 in $r_c^{min}~=~$1.4001 and for $\varphi_{2}$ we have $S_t^{min}~=~$12.9679 in $r_c^{min}~=~$1.6000~u.a.. For the Li$^{+}$, we have for $\varphi_{1}$ the value of $S_t^{min}~=~$12.9674 arise in $r_c^{min}~=~$0.9500~u.a. and for $\varphi_{2}$ we have $S_t^{min}~=~$12.9675 in $r_c^{min}~=~$1.1000~a.u.. So, the inclusion of the Coulomb correlation modifies, in some way, the balance between Coulomb and confining potential, modifying the confinement radius associated with the lowest entropy sum. Differences in $r_c^{min}$ due to improved model have a greater impact on H$^-$ anion, reinforcing its intriguing characteristics.   

In the context of the density functional theory~\cite{parr,NAGY19981} the informational study of confined two-electron systems has been done utilizing the one-electron density normalized to unity~\cite{entropiaatomosconfinados, majumdar}. Nevertheless, analysis using two-electron probability densities normalized to unity, as in the present work, are also presented in the literature involving free helium-like atoms~\cite{nassar4,NASSER20173892}. For more details on the different entropic formulations and how they can be contrasted see Ref.~\cite{guevara}. Lastly, we did not find $S_r$, $S_p$ and $S_t$ available entropies values in literature for confined helium-like atoms using two-electron probability densities normalized to unity and written in terms of two-electron wavefunctions to compare with our results.

\subsection{Strong confinement regime}{\label{Strong confinement}}

The strong or extreme confinement regime for confined atomic system can be defined when the influence of the confining potential becomes greater than the free atomic system potential to specific configurations (or $r_c$ values). One important characteristic of the strong confinement region is that the probability density is highly concentrated in position space, distinctly to weakly confined regions. The definition of specific radii for the strong confinement regime in two-electrons atoms can be based on the ionization radius~\cite{MONTGOMERY20102044,AQUINO2003326,FLORESRIVEROS20101246}. In our previous studies of confined harmonic oscillator~\cite{wallas_fred_educacional} and confined hydrogenic-like atoms~\cite{wallas_fred_internacional}, we have suggested that the entropy sum can be employed to determine a strong or rigorous confinement regime.

In the present work, based on the entropy sum we define three regions of confinement for the systems. In the strong confinement region, when $r_c \rightarrow 0$, the confinement potential is preponderant. On the other hand, the weak confinement correspond to the regions where $r_c \rightarrow \infty$ and the Coulomb potential is dominant. The intermediate confinement region is close of the value of $S_t^{min}$ where the effects of the confining and Coulomb potentials has approximately a similar impact or influence on the system (the confinement and Coulomb potentials are compensated).   

\begin{figure}[!ht]
\centering
\includegraphics[scale=0.55]{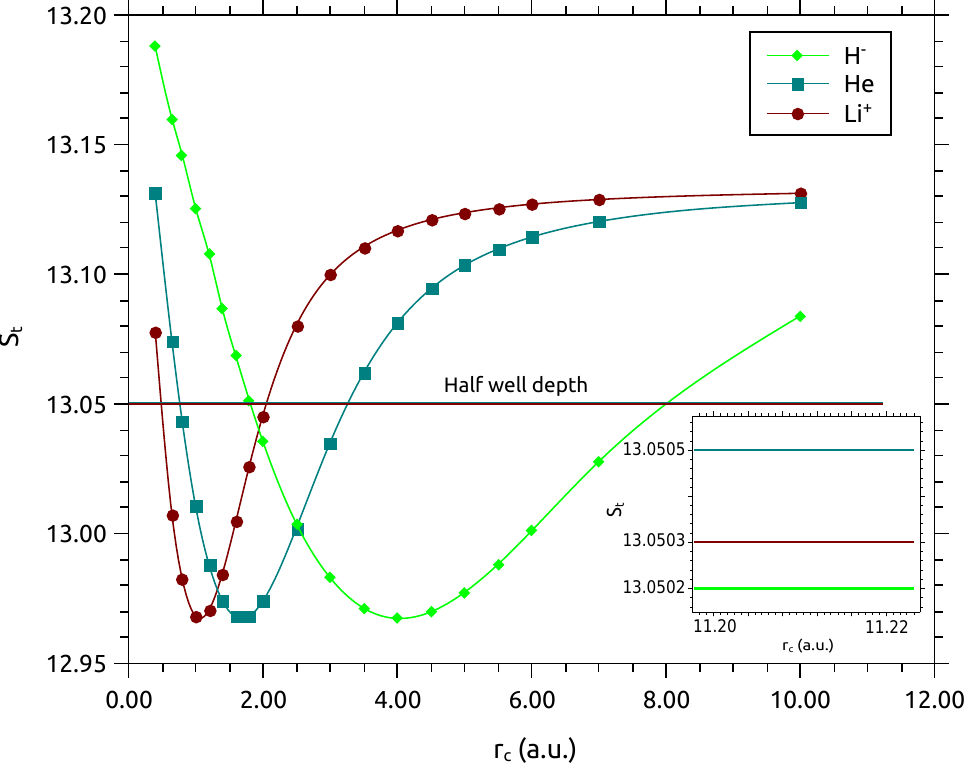}
\caption{Values of $S_t$ as a function of $r_c$ for the confined H$^{-}$, He and Li$^+$ atoms in their ground states. $S_t$ values using the $\varphi_2$ function. In the secondary graphic we highlight a portion in zoom of the main graph contrasting the emergence of the three horizontal lines.}
\label{somaentropica10}
\end{figure}

To illustrate and define more precisely the confinement regimes we use the values of $S_t$, obtained by $\varphi_2$ function, to plot the graph of the entropy sum for the H$^{-}$, He and Li$^+$ atoms in Fig.~\ref{somaentropica10}. The intermediate confinement region for each atom studied is defined between the values of $r_c$ where the horizontal line crosses the curve of $S_t$ versus $r_c$. The horizontal line marks the half well depth of the entropy curve, being given by   
\begin{equation}
\sigma_{[atoms]} = \frac{S_t [r_c=\infty] - S_t [r_c^{min}]}{2} \ ,  
\label{sigma}
\end{equation}
where the $S_t$ values are calculated in $r_c$ tending to infinity and in $r_c^{min}$ ($r_c$ value where $S_t^{min}$ occurs). The use of $\sigma_{[atoms]}$ provides a good approximation for the intermediate confinement. As defined by Eq.~(\ref{sigma}), the main graph of Fig.~\ref{somaentropica10} have horizontal lines referring to H$^{-}$, He and Li$^+$ atoms, however due to the graphic scale these lines overlap. In the secondary graphic of the same figure we highlight a portion in zoom of the main graph contrasting the emergence of the three horizontal lines.

In addition, in Fig.~\ref{somaentropica10} the strong and weak confinement correspond to the regions, respectively, left ($r_c \rightarrow 0$) and right ($r_c \rightarrow \infty$), of the intermediate confinement. Based on this analysis in Table~\ref{quadro} we organize the numerical values that limit the confinement regions and height of half well depth of the atoms studied.

\begin{table}[!ht]
\caption{Confinement region of the confined two-electrons atoms. Values of $r_c$ in atomic units system.}
\centering
\begin{tabular}{|c|c|c|c|c|}
\hline
Atoms   & height of half well depth & Strong       & Intermediate     & Weak                 \\ \hline
H$^{-}$ & 13.0502       & $r_c~<~$1.80  & 1.80$~\leq~r_c~\leq~$8.00 & $r_c~>~$8.00 \\ \hline
He      & 13.0505       & $r_c~<~$ 0.75 & 0.75$~\leq~r_c~\leq~$3.25 & $r_c~>~$3.25 \\ \hline
Li$^+$  & 13.0503       & $r_c~<~$ 0.50 & 0.50$~\leq~r_c~\leq~$2.10 & $r_c~>~$2.10 \\ \hline
\end{tabular}
\label{quadro}
\end{table}
Some of the main aspects of the strong or extreme confinement regime is the fact that, in this region, the effects of electronic correlation become essentially null, however the influence of the confining potential becomes greater than Coulomb potential. In this specific region, the system have  the probability density highly concentrated in position space~\cite{barbosa,prudente}. These characteristics are obtained in the present work through an elegant informational analysis.

\begin{figure}[!ht]
%\centering
\includegraphics[scale=1.5]{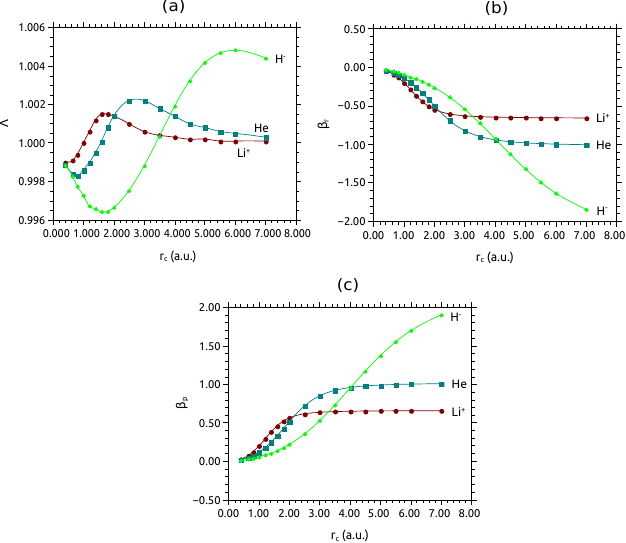}
\caption{Values of $\Lambda$, $\beta_r$ and $\beta_p$ as a function of $r_c$ for the confined H$^{-}$, He and Li$^+$ atoms in their ground states.}
\label{elementos}
\end{figure}

Initially, we examine the inclusion of the $V_{ee}$ term in Hamiltonian~(\ref{hamiltonianohelio}) through the quantity $\Lambda$, represented in terms of the entropy sums been determined by $\varphi_{1}$ and $\varphi_{2}$ functions, as follows:
\begin{equation}
\Lambda = \frac{S_t [\varphi_{1}]}{S_t[\varphi_{2}]} \ .  
\end{equation}
In addition, using the $S_r$ and $S_p$ entropies, computed also by $\varphi_{1}$ and $\varphi_{2}$, we have established the $\beta_r$ and $\beta_p$ quantities, that is, 
\begin{equation}
\beta_r = S_r \left[ \varphi_{1} \right] - S_r \left[ \varphi_{2} \right] 
\end{equation}
and
\begin{equation}
\beta_p = S_p \left[ \varphi_{1} \right] - S_p \left[ \varphi_{2} \right] \ .  
\end{equation}
In Figs.~\ref{elementos}, we present the curves of $\Lambda$, $\beta_r$ and $\beta_p$ as a function of $r_c$ for the examined atoms, respectively in graphs (a), (b) and (c). In situation (a) the $\Lambda$ values tend to unity when the $r_c$ values decrease. The graphs (b) and (c) indicate that the $\beta_r$ and $\beta_p$ values goes to zero when the confinement situation intensifies. These results imply that when $r_c$ goes to zero, the entropy values related to $\varphi_{2}$ function, in position space or in momentum space, tend to entropy values related to $\varphi_{1}$ function. Therefore, we conclude that the independent particle model can be successfully employed in the strong confinement region.

The values of the entropy sum of the investigated confined two-electrons atoms have interesting relation with our previous researches involving the confined hydrogenic-like atoms (He$^{+}$, H and Li$^{++}$) accessible in Ref.~\cite{wallas_fred_internacional}. For instance, for $r_c$~=~0.4000~a.u., we have the values of $S_t [\mathrm{H}^{-}]$~=~13.1878,  $S_t [\mathrm{He}]$~=~13.1312 and $S_t [\mathrm{Li}^{+}]$~=~13.0780. Moreover, 2$S_t [\mathrm{H}]$~=~13.1666,  2$S_t [\mathrm{He}^{+}]$~=~13.1072 and 2$S_t [\mathrm{Li}^{++}]$~=~13.0518. The numerical results show that the $S_t$ values of  H$^{-}$, He and Li$^{+}$ are approximately the double of H, He$^{+}$, and Li$^{++}$, respectively. This investigation denotes the loss of importance of the $V_{ee}$ term for small confinement radii, being in favor of the adoption of the independent particle model. 

\begin{figure}[!ht]
\centering
\includegraphics[scale=0.7]{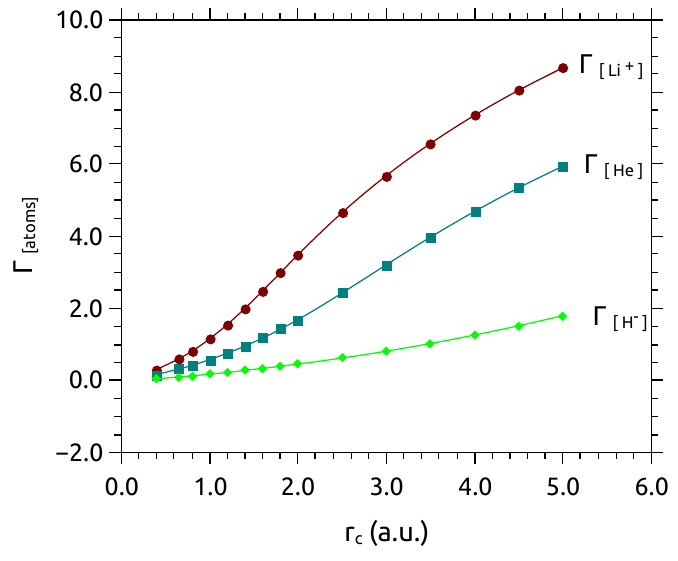}
\caption{Values of $\Gamma_{[atoms]}$ as a function of $r_c$ for the confined H$^{-}$, He and Li$^+$ atoms in their ground states.}
\label{t}
\end{figure}

We utilize the $2e^{-}$ system to auxiliary our investigation of the strong confinement region. This system is constituted by two non-interacting electrons confined by a impenetrable spherical cage (see Appendix). In this sense, we determine the quantity $\Gamma$ as follows:  
\begin{equation}
\Gamma_{[atoms]} =S_r [2e^-] - S_r[\mathrm{atoms}] \ .
\end{equation}
The $S_r [\mathrm{atoms}]$ values are obtained employing the $\varphi_{2}$ function, where the reported atoms are H$^{-}$, He and Li$^+$. In Fig.~\ref{t}, $\Gamma_{[atoms]}$ versus $r_c$ curves, we note that increasing the confinement effect the $\Gamma_{[atoms]}$ values go to zero. That is, the $S_r$ values of H$^-$, He and Li$^+$ tend for the results of $2e^{-}$, which leads us to establish that when $r_c$ go to zero the confining potential  becomes more relevant than the nuclear potential. Furthermore, the the effects of Coulomb correlation become negligible. These effects favor employment of the model of non-interacting electrons inside a sphere for the confined two-electrons atoms in the strong confinement regime.

The $S_r$ values obtained for the systems $2e^{-}$, H$^{-}$, He and Li$^+$ have the following relation: $S_r[2e^-] > S_r[\mathrm{H}^-] > S_r[\mathrm{He}] > S_r[\mathrm{Li}^+]$. Besides, in $r_c$~=~0.4000~a.u., we have the values approximately equal of $S_t [2e^{-}]$~=~13.2344, $S_t [\mathrm{H}^{-}]$~=~13.1878, $S_t [\mathrm{He}]$~=~13.1312 and $S_t [\mathrm{Li}^{+}]$~=~13.0780. Thus, we infer that in strong confinement the electrons are primarily bound to the confinement potential, instead of the nuclear potential, and the effects of electron-electron interaction lose importance. We found similar conclusion analyzing the Fig.~\ref{t}. For the $2e^{-}$ system the values of the entropic quantities and of the expectation value of the energy are available in Table~S5 of the Supporting Information. 

\section{Conclusion}{\label{conclusions}}

In this work, we determined the expectation values of the energy, $\langle E \rangle$, and the informational quantities $S_r$, $S_p$ and $S_t$, for the confined H$^{-}$, He and Li$^{+}$ atoms (ground states) confined at the centre of an impenetrable spherical cavity. The informational quantities were calculated using the two-electron probability densities normalized to unity. For this, we have employed three different variational wavefunctions.  

We obtained more accurate results of $\langle E \rangle$ using the wavefunctions that better take account of Coulomb correlation. The correlation term $\gamma(r_{12})$ goes losing importance when the confinement increases. The $\langle E \rangle$ values tend to the results of the unconfined system when $r_c$ goes to infinity. On the other hand, the energy values experience a considerable increase when $r_c$ value diminishes. The obtained $\langle E \rangle$ values have a good agreement with found results in literature.

The $S_r$ values increase, for the confined He and Li$^+$ atoms, using the $\varphi_1$, $\varphi_2$ and $\varphi_3$ wavefunctions, respectively. For H$^{-}$, the $S_r$ values increase of $\varphi_1$ to $\varphi_2$ and decreases of $\varphi_2$ to $\varphi_3$. This last effect occurs because in $\varphi_3$ we have a more localized probability density. The $S_p$ values decrease and the values of $r_c$, in which $S_t^{min}$ arise, is shifted in order of better take account of Coulomb correlation in the system. When $r_c$ value goes to infinity the $S_r$, $S_p$ and $S_t$ values go to the results of the unconfined system. The $S_p$ values using the function $\varphi_{3}$ have not been determined, fortunately this not produces changes on the main conclusions of our study. In the future we will return to this specific question.  

All the minimum values of $S_t$ obtained here satisfy the entropy uncertainty relation. We identified a situation of lower entropic uncertainty in $S_t^{min}$ where the Coulomb and the confining potentials are compensated. The symmetry $r_c^{mim}~\equiv~{3.0}/{Z}$~a.u  found for the two-electrons atoms using the independent particle model, $\varphi_{1}$, is broken with the incorporation of a part of the Coulomb correlation in $\varphi_{2}$ function.

Finally, based on the $S_t$ values, we define the regions of strong ($r_c \rightarrow 0$), intermediate (close of $S_t^{min}$) and weak confinement ($r_c \rightarrow \infty$) for the confined H$^{-}$, He and Li$^{+}$ atoms in their ground energy states. Analyzing the informational quantities, we concluded that in the strong confinement regime, the effects of Coulomb correlation become negligible. So, we can successfully employ the models of independent particle and two non-interacting electrons confined by a impenetrable spherical cage for the atoms of two electrons when the confinement effects are extreme. 

\begin{acknowledgements}
This work has been supported by the Brazilian agencies CAPES (Coordena\c{c}\~ao de Aperfei\c{c}oamento de Pessoal de N\'ivel Superior) and CNPq (Conselho Nacional de Desenvolvimento Cient\'ifico e Tecnol\'ogico) through grants to the authors. The authors thank the referees for careful reading of the manuscript and for helpful comments and suggestions.
\end{acknowledgements}

\begin{flushleft}
\textbf{Author contribution:} All authors contributed equally to the paper.
\end{flushleft}

\begin{flushleft}
\textbf{Supporting Information.} This manuscript contains supplementary information. \href{https://static-content.springer.com/esm/art%3A10.1140%2Fepjd%2Fs10053-021-00177-6/MediaObjects/10053_2021_177_MOESM1_ESM.pdf}{Click here to access.}
\end{flushleft}
\appendix

\section{}

The $2e^{-}$ system consists of two non-interacting electrons confined in an impenetrable spherical cage. The non-relativistic Hamiltonian in atomic units for one electron confined in an impenetrable spherical cage, $e^{-}$, is 
\begin{equation}
\hat{H} = -\frac{1}{2} \nabla^2 + V_c (r) \ .  
\label{hamiltonianop}
\end{equation}
The confining potential is defined as
\begin{equation}
V_c (r)=  \left \{ \begin{array}{ccc}
0 & \mathrm{for} &  0 < r < r_c  \\
\infty & \mathrm{for} &  r \geq r_c 
\end{array}\right. ,
\end{equation}  
being $r_c $ the confinement radius. The ground state wavefunction is 
\begin{equation}
\phi(r) = A\left[ \frac{\sin \left( \pi r / r_c \right) }{r}\right] \left( \frac{1}{4 \pi} \right) \ ,
\label{solgeralgaiolal0}
\end{equation}
where $A$ is a normalization constant. In this background, for the $2e^{-}$ system the expectation values of the energy, beyond the entropic quantities, are the double of the values obtained for the $e^{-}$ system. The $S_r$, $S_p$, $S_t$ and $\langle E \rangle$ values as a function of $r_c$ for the $2e^{-}$ system in ground state are available in Table~S5 of the Supplementary Material.



\bibliographystyle{spphys}       % APS-like style for physics
\bibliography{referencesepjd}   % name your BibTeX data base

\end{document}